\documentclass[%
 reprint,
 superscriptaddress,
 amsmath,amssymb,
 aip,
 jap,
]{revtex4-1}

\usepackage{graphicx}
\usepackage{dcolumn}
\usepackage{bm}
\usepackage{hyperref}
\usepackage[version=4]{mhchem}
\usepackage{xcolor} 

\begin{document}

\title{Modelling Grain Boundaries in Polycrystalline Halide Perovskite Solar Cells}

\author{Ji-Sang Park}
\email{jsparkphys@knu.ac.kr}
\affiliation{ Department of Physics, Kyungpook National University, Daegu, 41566, South Korea}

\author{Aron Walsh}
\email{a.walsh@imperial.ac.uk}
\affiliation{Department of Materials, Imperial College London, Exhibition Road, London SW7 2AZ, UK}
\affiliation{Department of Materials Science and Engineering, Yonsei University, Seoul 03722, Korea}

\date{\today}

\begin{abstract}
Solar cells are semiconductor devices that generate electricity through  charge generation upon illumination. For optimal device efficiency, the photo-generated carriers must reach the electrical contact layers before they  recombine. A deep understanding of the recombination process and transport behavior is essential to design better devices.  Halide perovskite solar cells are commonly made of a polycrystalline absorber layer, but there is no consensus on the nature and role of grain boundaries. This review paper concerns theoretical approaches for the investigation of extended defects. We introduce recent computational studies on grain boundaries, and their influence on point defect distributions, in halide perovskite solar cells. We conclude the paper with discussion of future research directions.
\end{abstract}

\keywords{halide perovskites, 
          grain boundaries,
          extended defects,
          first-principles,
          density functional theory
          }
\maketitle

\section{Introduction}

Perovskite solar cells have received a lot of attention partly because of the fast optimization of the device architecture and performance, which is illustrated in the rapid increase of the power conversion efficiency from 3.8 \% to 25.2 \%.\cite{kojima2009organometal,huang2017understanding}
Both inorganic (e.g. \ce{CsPbI3}) and hybrid organic-inorganic (e.g. \ce{CH3NH3PbI3}) materials have been studied.
The high performance of perovskite solar cells is due to the inherent material properties such as tunable band gap,\cite{ogomi2014ch3nh3sn,noh2013chemical} efficient charge generation, long diffusion length,\cite{stranks2013electron,shi2015low,tong2019carrier} and defect tolerance.\cite{steirer2016defect}
The solar cells are also made at relatively low temperatures,\cite{jeon2014solvent} leading to the production of high-quality solar cells at low cost.\cite{snaith2013perovskites}
Nowadays even higher efficiency has been achieved by perovskite/Si tandem solar cells,\cite{sahli2018fully}
and extensive efforts have been made to achieve large-scale solar cells with long term stability.\cite{li2018scalable,jung2019efficient,bai2019planar}

Perovskite solar cells are mainly made of polycrystalline materials, which means that a substantial amount of effort should have been devoted to understanding the effects of grain boundaries.\cite{huang2017understanding,wang2017scaling,lee2018role,tennyson2019heterogeneity,castro2019role,luo2019minimizing,han2019interface,chen2019causes} 
Grain boundaries are known to affect a variety of physical, chemical, and material properties, such as recombination, transport, and even degradation; however, our general knowledge of grain boundaries in halide perovskites remains far from complete. 
In this review, we focus our scope on the electrical and optical properties of the grain boundaries as there are many unanswered questions to be solved. These include the nature of nonradiative electron-hole recombination in halide perovskites.\cite{stranks2017nonradiative} 
Since there a range of terminology frequently used in texts without detailed explanation, we start from the basics of the grain boundary in crystals and studies in other inorganic materials.
We not only outline our current understanding of grain boundaries in halide perovskites but also discuss the other extended defects and physical properties that need to be addressed in the future studies.

\section{Fundamentals of grain boundaries}

\begin{figure}
\includegraphics[width=0.5\textwidth]{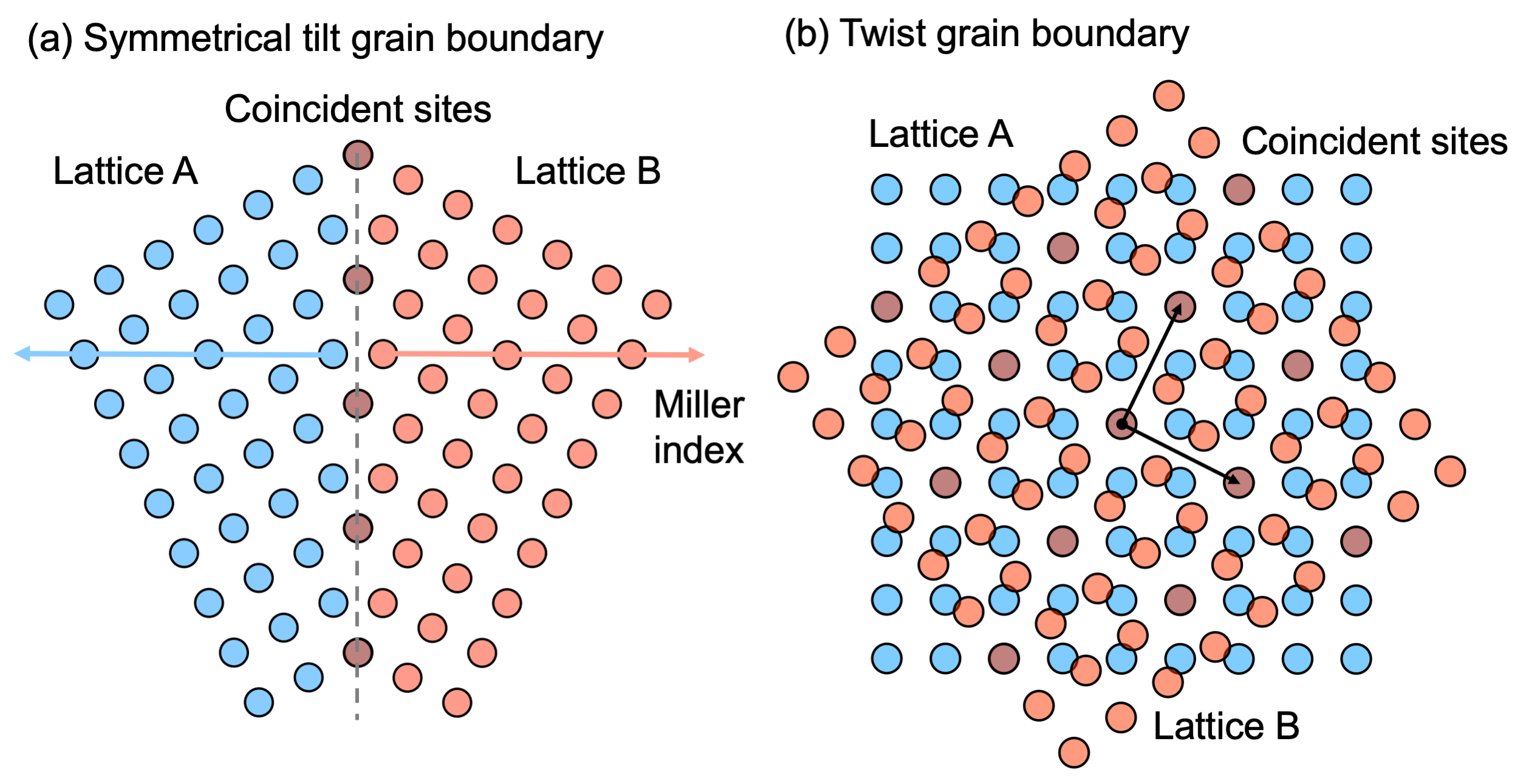}
\caption{\label{fig:1} Illustration of $\Sigma$5 (120) grain boundaries. 
(a) A symmetrical tilt grain boundary, and (b) a twist grain boundary. 
In (a), the boundary plane is denoted by a dashed line. In (b), the boundary plane is in between the two overlapped planes. Circles with different colors represent the lattice points of grains. When two neighboring lattices are expanded to the other side of the boundary, one of every five lattice points overlap, resulting in the $\Sigma$ value of 5.
} 
\end{figure}

Polycrystalline materials are composed of randomly oriented grains.
Grain boundaries are boundaries between such grains, and are typically two-dimensional.\cite{sutton2006interfaces} 
Grain boundaries can be categorized by the Miller indices of the grains and the rotation angle. 
For instance, symmetrical tilt grain boundaries, which are also known as twin boundaries, are formed by two grains with the equivalent Miller index and the zero rotation angle. 
On the other hand, a twist grain boundary is characterized by a non-zero rotation angle when the rotation axis is perpendicular to the boundary.
A characteristic parameter widely used is the $\Sigma$ value,\cite{cai2016imperfections} which represents how much the two neighboring grains share coincident sites accross the lattice. 
Perfect materials are considered to have the $\Sigma$ value 1, and larger value indicates that fewer coincident sites form at the grain boundary.
Grain boundaries can have one-dimensional or two-dimensional order in their atomic structure.\cite{cai2016imperfections,yin2019ceramic}

Because every material is polycrystalline in macroscopic quantities, the role of grain boundaries on the material properties has been investigated in many classes of materials.
Grain boundaries have been subject of interest in metallurgy for a long time because mechanical properties of metals are highly correlated with the density and distribution of grain boundaries.\cite{lu2016stabilizing}
In the community of thin-film solar cells, there is also growing evidence that grain boundaries can be made beneficial for transport properties. 
One of the well-known examples is superior photo-conversion efficiencies of polycrystalline CdTe solar cells compared to crystalline CdTe.\cite{visoly2006understanding,li2013carrier}
To explain this counter-intuitive result, grain boundaries in CdTe have been discussed as being beneficial.\cite{visoly2004polycrystalline} 
One hypothesis is that Cl impurities are segregated at grain boundaries, which results in local p-n junctions, resulting in better separation of charge carriers and reduced recombination.\cite{li2014grain}
Similarly, attempts have been made in other materials to create local p-n junctions by inverting the charge carriers of grain boundaries with respect to grain interiors.\cite{chen2018efficiency,xu2019defect} 
Besides the benefits on the electrical properties, impurities segregated at grain boundaries might form precipitates, which can lead to a lower impurity concentration in the grain interior, promoting the gettering.\cite{lu2003effects}

Some studies show that grain boundaries can be relatively benign even though the atomic structure is far from the crystalline order. For instance, grain boundaries and dislocations in Si are relatively benign partly because the over-coordinated Si atoms at the grain boundaries do not introduce deep gap states.\cite{kohyama1988atomic,chelikowsky198230}

Grain boundaries, however, are generally thought to be detrimental for device performance because of faster carrier recombination and adverse band edge positions.\cite{klie2005enhanced,kuo2018grain} 
For instance, first-principles calculations show that oxygen vacancies can be generated more at grain boundaries in \ce{YBa_2Cu_3O_{7-$\delta$}} (YBCO) because of the inherent strain, resulting in the lower hole concentration.\cite{klie2005enhanced} 
Other first-principles calculations have also shown that some grain boundaries in CdTe, without impurities, can introduce deep levels in the band gap.\cite{yan2003structure,park2015stability}
These extended defects can be passivated partly by impurities or isovalent element substitution.\cite{park2016effect}
Although there are some examples of beneficial grain boundaries, generally we should expect them to act as recombination centers in solar cells and therefore hamper charge extraction, unless specific passivation routes have been identified and applied.\cite{moseley2015recombination}

\section{Models to investigate grain boundaries}

\subsection{Non-atomistic models}

\begin{figure}
\includegraphics[width=0.45\textwidth]{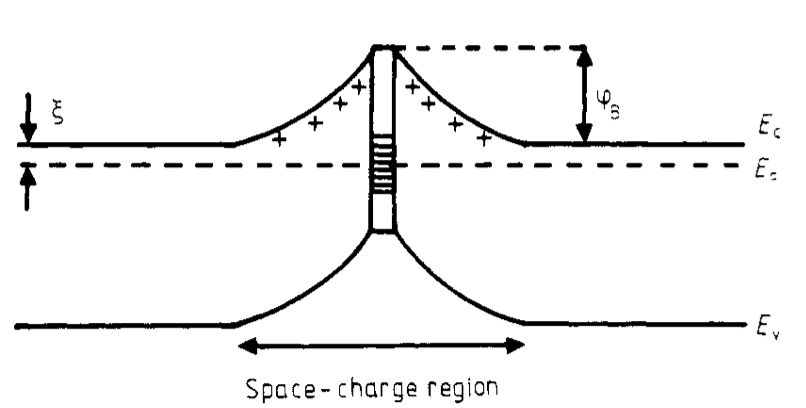}
\caption{\label{fig:2} 
A schematic one-dimensional band diagram of a grain boundary in an n-type semiconductor. The space-charge region is formed due to grain boundary states.
Figure is adapted with permission from Reference 42.} 
\end{figure}

\textit{Stability.} 
Read and Shockley derived a phenomenological function that describes interfacial energy assuming that grain boundaries consist of dislocations.\cite{read1950dislocation} Their model indeed described the energy of grain boundaries with small misorientation angle (also known as low angle grain boundaries) well. However, it could not describe the energy of the high angle grain boundaries and the existence of local minima. 

\textit{Defect segregation.}
Grain boundaries are known to serve as reservoirs for point defect (e.g. vacancy, interstitial or substitutional impurity) segregation. 
This behavior is generally understood in terms of two contributions: elastic and electrostatic.\cite{sutton2006interfaces,kliewer1965space,desu1990interfacial,gregori2017ion}
Elastic interactions between the defects and grain boundaries can be understood as follows.
If an impurity atom replaces a host atom, substitutional defects are formed and will generate stress that is proportional to the atomic size mistmatch.
Grain boundaries also likely to generate pressure in their vicinity because of different atomic number density and structure compared to the perfect crystal.
Electrostatic interactions can dominate when charged defects are formed.
The distribution of charges and defects can be obtained through consideration of long-range electrostatics (i.e. Poisson's equation).

\textit{Transport properties.} 
The function of solar cells is to extract  charges generated by absorbing light into electrical contacts, and in this regard, the transport properties are of particular interest.
In polycrystalline semiconductors, grain boundaries are expected to have deep trap states because of incomplete chemical bonding at the boundaries and their role as reservoirs for point defect segregation.
If there is no band bending near grain boundaries, defects will start to trap free carriers, and as a result, a potential energy barrier is built that eventually inhibits transport of charge carriers from grain to grain.
Several theories have been developed to explain the transport behavior of grain boundaries.\cite{seto1975electrical,landsberg1984effects,card1977electronic,nelson2003physics}
Those have successfully shown that the barrier height increases with the trap density at the grain boundaries as the space charge is increased. 
This results in reduced conductivity and increased grain boundary recombination.



\textit{Recombination.} 
The non-radiative recombination rate of a solar cell can be described by Shockley-Read-Hall (SRH) recombination statistics.\cite{oualid1984influence,edmiston1996improved} 
Assuming a single grain boundary trap level in the gap, the SRH recombination rate under steady-state non-equilibrium condition can be represented in terms of the surface recombination velocity:
\begin{equation}
    R_{\mathrm{SRH} } = \frac{S_n S_p (np - n_i^2)} {S_n (n + n_t) + S_p (p + p_t)},
\end{equation}
where $S_n$ and $S_p$ are the electron and hole recombination velocities.
$n_t$ and $p_t$ are $n_i \exp{(E_t-E_i)/k_B T}$ and $n_i \exp{(E_i-E_t)/k_B T}$, respectively. 
$n_i$ is intrinsic carrier density, $k_B$ is the Boltzmann factor, and $T$ is temperature.
$E_t$ is the trap level, and $E_i$ represents the intrinsic Fermi level.
It has recently become possible to calculate the SRH rate arising from equilibrium populations of point defects from first-principles.\cite{kim2020upper}

\subsection{Atomistic simulations}

In the 1970s, several methods were developed to calculate the grain boundary energy using interatomic potentials.\cite{hasson1970structure,weins1972structure} 
These attempts are clearly different from  previous phenomenological models because we can search the atomic configuration space directly. 
Stable configurations can be searched by minimization of the grain boundary energy.
Then the grain boundary energy was calculated as a function of the misorientation angle, and found to be effective to overcome the previous problems of phenomenological models.\cite{read1950dislocation}
Simple inter-atomic potentials such as Morse and Lennard-Jones potentials were used in early studies,
but more sophisticated potentials are currently used.\cite{olmsted2009survey,holm2010comparing,restrepo2013genetic,ratanaphan2015grain} 

The above approach based on structure searches using interatomic potentials were successful to predict the grain boundary atomic structure in metals; however, there was a need for a quantum mechanical description of semiconductors.
Tight-binding methods were adapted to understand extended defects, and the density of states (DOS) of grain boundaries was calculated as well.\cite{de1980electronic,thomson1984theoretical,chadi1985new,paxton1988simple}
In 1986, when a first-principles method was first applied to study twin boundaries in crystals,
empirical tight binding methods were employed to optimize the structures of grain boundaries in Si because of the lower computational cost.\cite{divincenzo1986electronic}
More recently, an effective tight-binding model was developed to understand a grain boundary in $\mathrm{YBa_2Cu_3O_{7-\delta}}$ superconductor.\cite{graser2010grain}
Motion and annihilation of grain boundaries in graphene has been investigated using a molecular dynamics tight-binding method as well.\cite{lee2013atomistic}

\subsection{First-principles simulations}
To fully describe the stability and the electronic structure of materials, a fully quantum mechanical calculation method without empirical parameters is ideal.
First-principles density functional theory (DFT) meet these needs\cite{kohn1999nobel} and can be used to investigate the stability and the electronic structure of grain boundaries.
We note that there a number of technical challenges for halide perovskites owing to strong relativistic effects (due to Pb) and dynamic structural effects.\cite{whalley2017perspective}

\textit{Stability.} Since periodic boundary conditions are typically employed in simulations of crystals, a supercell model may contain two interfaces if there is no vacuum region in the supercell. 
Since a grain boundary is a type of interface, the method used to obtain the interface energy can be directly applicable:\cite{park2018quick}
\begin{equation}
E_f(\mathrm{GB}1) + E_f(\mathrm{GB}2) = (E_{tot}(\mathrm{GB})-\Sigma_i n_i \mu_i) / A,
\end{equation}
where $E_f(\mathrm{GB})$ is the formation energy of a grain boundary, $E_{tot}(\mathrm{GB})$ is the total energy of a given supercell with two grain boundaries. $n_i$ and $\mu_i$ are the number of atom of atomic species $i$ and the corresponding chemical potential. 
$A$ is the area of the grain boundary in the supercell. If the two interfaces are exactly the same, the formation energy becomes
\begin{equation}
E_f(\mathrm{GB}) = (E_{tot}(\mathrm{GB})-\Sigma_i n_i \mu_i) / 2A.
\end{equation}
In many cases, grain boundaries in the supercells are not identical, and therefore charges can be transferred between the grain boundaries and affect the formation energy. To obtain the formation energy of a single grain boundary, we need to employ a slab geometry that contains one interface and two surfaces. As there are two surfaces, their contribution to the formation energy should be subtracted.
Park \textit{et al.} used slab geometry and successfully obtained the formation energy of grain boundaries in CdTe.\cite{park2015stability}

\textit{Electronic structure.} 
The electronic structure of grain boundaries is often calculated with DFT using the generalized gradient approximation (GGA) exchange-correlation functional, which underestimate the band gap.\cite{perdew1985density}
Hybrid DFT calculations,\cite{heyd2003hybrid,heyd2005energy} which are commonly used to correct the band gap of semiconductor materials nowadays, are currently too computationally heavy for describing grain boundaries. 
Often-used strategies are to introduce the on-site Coulomb interaction\cite{yin2013defect,yin2014engineering} or hybrid calculations only for analysis of the electronic structure.\cite{park2015stability,park2016effect,park2019stabilization}
To further reduce the computational cost, a sparse \textit{k}-point grid mesh can be used for the Fock exchange potential or non-self-consistent-field calculations can be performed.\cite{pan2018spin,park2019stabilization,park2020examination}

\textit{Prediction of atomic structure.} 
An important question is how to generate a representative three-dimensional atomistic model of a grain boundary. 
Structural properties of grain boundaries can be identified by electron backscattering diffraction (EBSD) at a microscopic scale.\cite{humphreys2001review,moseley2015recombination} 
Typically various types of grain boundaries are observed. Further atomistic details can be obtained by transmission electron microscopy (TEM) measurements.\cite{yan2003structure,li2014grain,liebscher2018strain} 

A potential problem, however, is that information gathered from the experiment like two-dimensional images could be insufficient to construct a three-dimensional atomic structure. 
We also have few images than grain boundaries formed in real samples.
To overcome this problem, statistical techniques such as genetic algorithms have been developed.
Grain boundaries in metals have been investigated using the force field calculations, which are relatively cheaper than DFT calculations.\cite{olmsted2009survey,restrepo2013genetic}
Grain boundaries in semiconductors, on the other hand, are better to be investigated by the quantum mechanics code due to the importance of the electronic structure.
Chua \textit{et al.} investigated both stoichiometric and non-stoichiometric grain boundaries in \ce{SrTiO3}.\cite{chua2010genetic}
In their framework, thousands of trial configurations were explored using empirical interatomic potentials, and thereafter structures were refined using first-principles electronic structure methods.
Similarly Park \textit{et al.} performed DFT calculations but using the atomic orbital basis to explore the configuration space.\cite{park2019stabilization} 
Some screened structures were re-examined using plane-wave basis methods.
We also note that the mirror symmetry of symmetrical tilt grain boundaries in semiconductors can be broken as a result of the rigid body translation as examined in the literature recently.\cite{marquis2004finite,liebscher2018strain}

\section{Experimental findings}

\textit{Beneficial grain boundaries.}
The first question to be answered is whether grain boundaries in halide perovskites are beneficial or not from the device perspective.
Early studies using Kelvin probe force microscopy (KPFM) and conductive atomic force microscopy (AFM) reported that grain boundaries are beneficial because charges are efficiently separated and collected through grain boundaries.\cite{yun2015benefit,kim2015efficient,li2015microscopic}
Later Yun \textit{et al.} used KPFM to detect local surface potentials caused by ion profiles in halide perovskites.\cite{yun2016critical} 
Their KPFM experiments have shown that the contact potentials difference (CPD) of grain boundaries and grain interior exhibit different trends. The grain boundary always had a lower CPD value than grain interior when there is no bias voltage. 
However, applying positive bias (more than 1 V) makes the grain boundaries to have higher CPD than the grain interiors, whereas negative bias exhibit the opposite effect. It also took several minutes for the CPD value to return to its initial value after the bias voltage is removed. 
Based on these results, the authors conclude that there were more ions at the grain boundary initially or ions migrate easily through the grain boundaries.  
A phenomenological model developed by the authors claims that redistribution of ions under illumination condition results in stronger band bending at grain boundaries.
The contact potential difference at grain boundaries in KPFM measurements was found to be modulated by additives.\cite{faraji2018grain}

\textit{Neutral grain boundaries.}
Some studies focused on the transport properties of grain boundaries.
MacDonald \textit{et al.} found that grain boundaries are electrically resistive, at least near the top of the film.\cite{macdonald2016methylammonium}
Reid \textit{et al.} observed that mobility-yield products decrease with decreasing the grain size.\cite{reid2016grain}
Yang \textit{et al.} constructed a kinetic model of charge transport and recombination process based on their high-resolution confocal fluorescence lifetime imaging microscopy experiments,\cite{yang2017grain} 
and pointed out that the weaker PL intensity does not necessarily mean a shorter lifetime of carriers. 
Snaider \textit{et al.} also concluded that the carrier transport is slowed down by grain boundaries.\cite{snaider2018ultrafast}
It was also discussed that long carrier lifetime can compensate for the higher resistivity at the grain boundary.
Sherkar \textit{et al.} performed device simulation modeling and found that grain boundaries become relatively inert when the charged traps become neutral after charge trapping.
\cite{sherkar2017recombination}

\textit{Detrimental grain boundaries.}
Local fluorescence lifetime imaging experiments have shown that the photoluminescence intensity is lower near the grain boundary than the center of the grain in methylammonium lead iodide (\ce{CH3NH3PbI3}).\cite{de2015impact} 
This result indicates that grain boundaries are active for non-radiative recombination. 
Passivation of the boundaries (e.g. using pyridine) resulted in brighter PL.
Scanning Electron Microscopy (SEM), which allows us to study surface morphology, is not sufficient to identify crystallographic information of grains and grain boundaries.
Electron backscatter diffraction (EBSD) is the standard method to measure crystallographic information of grains, but its usage was hampered because of beam damage to halide perovskite samples. Adhyaksa \textit{et al.} used a solid-state EBSD detector with better sensitivity to resolve this problem and found that grain boundaries in halide perovskites can act as recombination centers.\cite{adhyaksa2018understanding}

\begin{figure}
\includegraphics[width=0.45\textwidth]{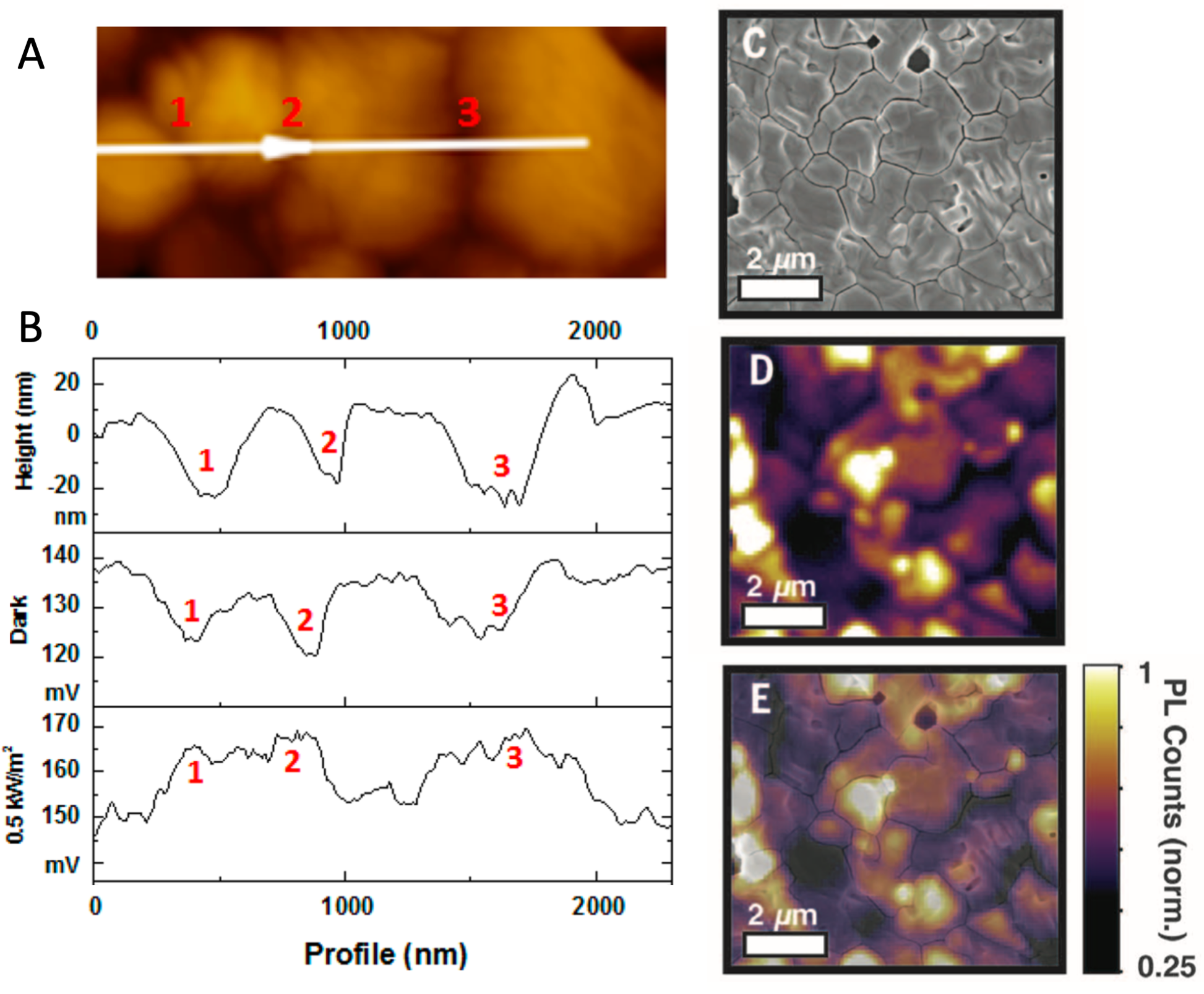}
\caption{\label{fig:3} 
(A) Topography map and (b) line profile data of topography and contact potential difference (CPD) under different conditions.
(C) SEM, (D) fluorescence spectroscopy and (E) their composite image, showing that photoluminescence intensity spatially varies.
Figures adapted with permission from Ref. 82 and 92.} 
\end{figure}

\section{First-principles studies}

\subsection{Neutral grain boundaries}
Yin \textit{et al.} studied two kinds of GBs, $\Sigma$3 (111) GB and $\Sigma$5 (310) GB in \ce{CH3NH3PbI3}.\cite{yin2015origin} In their DFT-PBE calculations, they found that grain boundary models do not introduce deep levels in the band gap even though there are I-I bonds formed, which are not formed in perfect \ce{CH3NH3PbI3}, as well as Pb dangling bonds. 
\cite{yin2014unusual}
This is in line with the fact that iodine vacancy (Pb dangling bonds) and iodine interstitials, which form I$-$I bonds, are shallow defects in their previous study.\cite{yin2014unusual} 
Iodine anti-site defects also form I$-$I bonds and even introduce deep levels in the band gap, but those were not as stable as I interstitials. 
Besides these defects, Pb antisite defects created deep levels in their PBE calculations without spin-orbit coupling, and all of them had relatively high formation energy.
The electronic structure of $\Sigma$3 (111) GB was more carefully examined by using the hybrid functional with spin-orbit coupling, but they were not able to find a deep level in the gap. 
They ascribed the origin of the deep-state-free GBs in \ce{CH3NH3PbI3} is due to the strong \textit{sp} coupling of the valence band maximum and to the large atomic size of \ce{CH3NH3PbI3}. 
The former and the latter results in the higher band edge and the shallower defect states. 
Extrinsic elements such as Cl and O were found to be stable at the grain boundaries, and weaken the halogen-halogen bonds (i.e. I$-$I) at GBs and thus are able to reduce the density of shallow trap states.

Guo \textit{et al.} performed more comprehensive studies on the grain boundaries in halide perovskites (\ce{CsPbX3} where X = I, Br, and Cl).\cite{guo2017structural} 
Using DFT, they investigated symmetrical tilt grain boundaries having four Miller indices. 
Remarkably, they considered rigid body translation in the simulation to find stable geometry of the grain boundaries. 
Contour maps of the grain boundary energies were also reported. The grain boundary energies were obtained and based on those data, some stable structures were selected. Electronic structure calculations, performed  using DFT-PBE, showed that the stable structures do not have any deep gap states, consistent with the previous study.\cite{yin2015origin} 
The electronic structure of \ce{CH3NH3PbI3} was also examined using the same geometry, but it also had no deep gap states.

\begin{figure}
\includegraphics[width=0.4\textwidth]{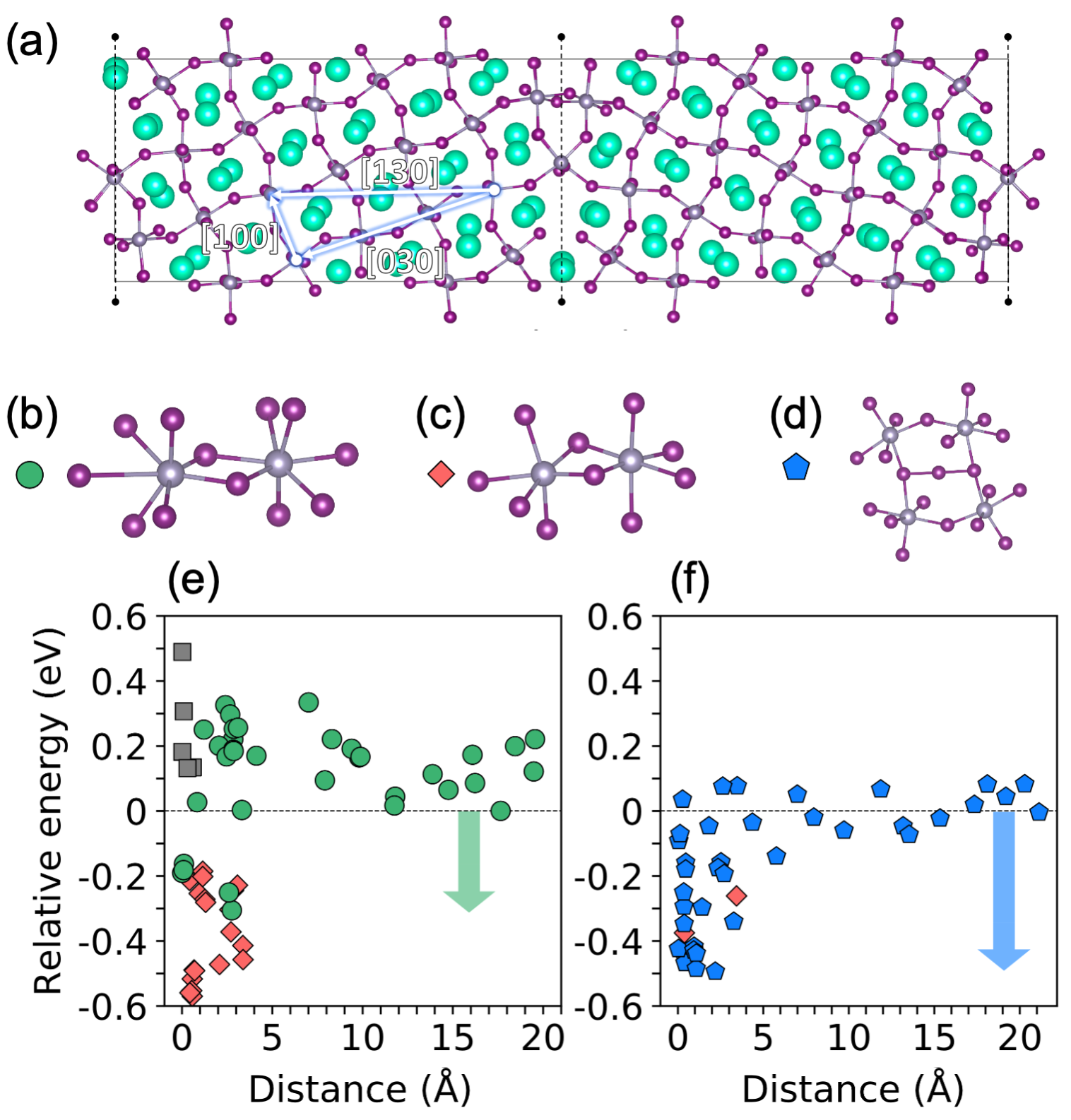}
\caption{\label{fig:4} (a) Atomic structure of a $\Sigma$5 [130] grain boundary in \ce{CsPbI3}. The boundary of the cells is represented by the solid lines. 
The dashed lines in the middle represent the grain boundary. (b) Split-interstitial configuration of iodine interstitial (I$_i$), denoted by a green circle.
(c) I$_i$ passivating under-coordinated Pb atoms, denoted by a blue pentagon. 
(d) I$_i$ with an iodine trimer (I-I-I) denoted by an orange rhombus. 
(e,f) The relative formation energy of I$_i$ in 1$-$ and 1$+$ charge state, respectively, as a function of the distance from the grain boundary. 
The results show the segregation of I$_i$ defects to the grain boundary. Figures adapted with permission from Ref. 97.} 
\end{figure}

\subsection{Defect-mediated recombination}
The atomic structure of grain boundaries differ from the bulk region, and thus defect properties can be affected.
Thind \textit{et al.} studied the grain boundaries and other planar defects that can be formed in \ce{CsPbBr3}.\cite{thind2019atomic} They first made \ce{CsPbBr3} nanocrystals and then fused them to make larger crystals. Various boundaries can be generated depending on how the nanocrystals are aligned. Based on the atomic structure observed experimentally, they constructed an atomic structure model for DFT calculations and investigated the electronic structure. 
In their study, grain boundaries cause band offsets and impact electron transport. 
A specific type of grain boundary ($\Sigma 5$) repels electrons and attracted holes. 
However, Ruddlesden-Popper faults repel both kinds of carriers. 
This means that the transport and optoelectronic properties of grain boundaries are greatly influenced by the atomic structure of the boundary.
Interestingly, their calculations predict that the bromine vacancy could cause relatively deep levels.\cite{thind2019atomic}  
It is worth pointing that PBE  describes defect properties of \ce{CH3NH3PbI3} quite differently to hybrid functionals,\cite{du2015density,meggiolaro2018iodine} which could impact the conclusions.

In an early study done by Shan \textit{et al.}, intrinsic defects were found to segregate to boundaries.\cite{shan2017segregation}
Since they performed the calculation using PBE with spin-orbit-coupling, the band gap was underestimated and only anti-site defects were assigned to be deep traps.
Later Park \textit{et al.} re-visited iodine interstitial defects,\cite{park2019accumulation} which introduce deep levels in the band gap\cite{du2015density,whalley2017h,meggiolaro2018iodine} and diffuse fast.\cite{yang2016fast,futscher2019quantification}
Iodine interstitials were found to easily segregate at the grain boundary, whichever charge state it has. 
The driving force of the segregation has been attributed to the structural relaxation, which is parameterized with the distance between iodine atoms forming the interstitial defect.
The results can be understood as the lower atomic density at the grain boundaries promote room for relaxation and hence energy lowering.
The numerical solution of Poisson's equation revealed that both donor and acceptor defects are heavily compensated at the grain boundaries. To investigate the effect of the environment on the defect levels, Park \textit{et al.} assumed halide dimers and trimers embedded in a dielectric medium and found that the acceptor (I$_i^{1-}$) is expected to be shallower and the donor state  (I$_i^{1+}$) deeper. 
The high concentration of deep traps can shorten the carrier lifetime through defect-assisted recombination at grain boundaries.

Meggiolaro \textit{et al.} also performed first-principles calculations to investigate the effect of environment on the formation energy of iodine interstitial defects.\cite{meggiolaro2019formation} They found that the defect formation energy at the surface was significantly lowered compared to bulk. 
Based on these results, they constructed a phenomenological equation to estimate defect formation energy as a function of grain size, which corresponds to the weighted average of defect formation energies corresponding to the bulk and surface. Simulation results showed that the more defects are easily formed as the grain size decreases.

We note that Hentz \textit{et al.} developed an experimental setup to measure the photoluminescence of laterally biased sample and concluded that nonradiative recombination centres migrate through grain boundaries.\cite{hentz2019visualizing}
Among several potential defects, iodine interstitials were discussed to be the best candidate to explain the result.
This is also consistent not only with the recent DFT calculation results that nonradiative iodine interstitials defects are easily accumulated at the grain boundaries\cite{park2019accumulation},  but also with the previous experimental results of fast ion migration through grain boundaries.\cite{shao2016grain,xing2016ultrafast}

\subsection{Band gap narrowing}
Although many computational studies overlooked anion mixing, McKenna has shown that the halide composition ratio can vary spatially.\cite{mckenna2018electronic} According to his first-principles calculation, the 
\{111\} twin boundary in pure formamidinium lead iodide only creates a small barrier of less than 0.1 eV. However, in the mixed-cation mixed-halide perovskite, Cs and I atoms were segregated at the twin boundary. The I accumulation caused the higher valence band edge at the boundary by more than 0.2 eV than in the bulk region, indicating that the photo-generated carriers could be recombined at the twin boundary.

Long \textit{et al.} performed molecular dynamics simulations and found that a grain boundary in pure \ce{CH3NH3PbI3} has a higher valence band edge than the bulk region.\cite{long2016unravelling} 
In their study, the reduced band gap and the higher coupling between the band edges result in the faster electron-hole recombination at grain boundaries. 
Cl incorporation reduced the coupling and thus the recombination became weaker.

\subsection{Passivation strategies}

If grain boundaries act as nonradiative recombination centers, then the origin of the deep levels should be identified, removed or passivated. 
Considering their importance for device efficiency and possibly lifetime, various attempts have been made to passivate the grain boundaries.\cite{lee2018role,castro2019role,chen2019causes,luo2019minimizing} 
Here, we introduce some studies showing consistency with DFT calculations.
On the experimental side, compositional engineering is a well-known method to enhance device efficiency.\cite{stranks2013electron,jeon2015compositional}
de Quilettes \textit{et al.} found a positive correlation between the PL intensity and Cl composition by using energy dispersive x-ray spectroscopy with confocal fluorescence maps.
Zheng \textit{et al.} employed a surface model and claimed that Cl can passivate ionic point defects (e.g. Pb$_\mathrm{I}$ anti-site) accumulated at the surface, noting that the major defects at the surface were uncertain at the moment of study.\cite{zheng2017defect}
On the computational side, Meggiolaro \textit{et al.} has found that Br interstitials and Cl interstitials introduce shallower acceptor levels than I interstitials.\cite{meggiolaro2018iodine}
Cl incorporation at the grain boundaries can be effective in this regard as the deeper I defects are replaced by shallower Cl defects.

Another category is the passivation of surface defects by extrinsic impurities or molecules.
For instance, Noel \textit{et al}. found that Lewis bases such as thiophene and pyridine can be used to reduce nonradiative recombination in halide perovskites.\cite{noel2014enhanced} 
They suggested that the molecules can be bound to defects (vacancies) on surfaces or grain boundaries, passivating defects and improving performance accordingly.
Later Shao \textit{et al.} have claimed that PCBM molecules can also passivate grain boundaries based on experimental data,\cite{shao2014origin}
and later Xu \textit{et al.} also found the same conclusion based on collaboration between experiment and modelling.\cite{xu2015perovskite}
In their DFT calculation, PCBM adsorption passivates the grain boundaries by making the deep levels of I$_{\mathrm{Pb}}$ closer to the conduction band minimum.

\section{Remaining open questions}

We have discussed how first-principles methods have been used to describe the structure and properties of grain boundaries in halide perovskites.
Here, we highlight some of the open issues in the topic.

\textit{Twin domains.} 
The formation of twin domains in \ce{CH3NH3PbI3} has been reported based on TEM and selected area electron diffraction (SAED) experiments.\cite{rothmann2017direct} 
In the TEM experiments, the striped contrast patterns (alternating bright and dark colors) were observed. 
Also in the SAED experiments, the split spots were observed in the striped domains. 
Morphology, however, was found to be not correlated with stripe contrast.
It has been claimed in another study that twin defects lowers the solar conversion efficiency, which the absorption coefficient were not affected.\cite{Tan2019} 
The formation of twin boundaries was measured from the shift of the (100) \textit{d} peak in TEM measurements.  

\textit{Mixed phases.} 
There is growing evidence that halide perovskites are not a single phase in real samples.
Kim \textit{et al.} have reported that tetragonal and cubic \ce{CH3NH3PbI3} can coexist at room temperature.\cite{kim2018self} 
They also observed superlattices composed of cubic and tetragonal phases in their TEM analysis. 
As there is no compositional change in their analysis, the superlattices were concluded to be formed as a result of intrinsic structural changes.
The detailed formation mechanism, however, and their effects on the device performance are not clearly revealed by first-principles calculations.

\textit{Internal grain structure.}
Using photoluminescence microscopy, Li \textit{et al.} have reported the formation of subgrain boundaries which cannot be observed by conventional Atomic Force Microscopy (AFM) and SEM measurements.\cite{li2018subgrain}
Those boundaries were reported to act as non-radiative recombination centers and also restrict carrier diffusion.
Jones \textit{et al.} used synchrotron scanning micro-XRD measurements
with local time-resolved PL measurements to identify that lattice strain is directly associated with enhanced defect concentration and therefore non-radiative recombination.\cite{jones2019lattice}
Jariwala \textit{et al.} have reported that even local orientation may vary even inside a grain, exhibiting higher recombination.\cite{jariwala2019imaging}
This finding is in contrast to a common belief that materials are aligned in a certain direction in a grain.

\textit{Dynamic properties.}
Most of the studies investigated the defects in temporal or spatial average, however, time-dependent phenomena should be investigated to obtain a complete picture of the grain boundary.
On experimental sides, Snaider \textit{et al.} investigated carrier transport phenomena using Transient Absorption Microscopy (TAM).\cite{snaider2018ultrafast}
Later Jiang \textit{et al.} investigated carrier dynamics using SEM correlated to TAM.\cite{jiang2019transient}
The latter study found that grain boundaries have an increased population of the sub-band-gap states than grain interior, higher quasi-Fermi energy, and faster carrier cooling rate.
The origin of the shallow state was suggested to be I$-$I bonds at the grain boundaries, partly based on a previous DFT calculation.\cite{yin2015origin}
Certainly, future studies should account for the dynamics of the photo-generated carriers.

\section{Outlook}

We have outlined several ways to investigate grain boundaries. Early studies employed phenomenological non-atomistic methods, but the development of computer simulation methodologies and the high-performance computers have allowed us to study grain boundaries using first-principles materials modelling. There is an urgent need to study various extended defects that can be generated in halide perovskite using this methodology. 
Trying to narrow the gap between the calculations and experiments should be pursued as well. For instance, various techniques being developed in point defect studies should be introduced to study to extended defects that are ubiquitous in the polycrystalline thin films being used in solar cells.

\acknowledgments
This work was supported by a National Research Foundation of Korea (NRF) grant funded by the Korean government (MSIT) (No. 2018R1C1B6008728).
This work was supported by a National Research Foundation of Korea (NRF) grant funded by the Korea government (MSIT) (No. 2019M3D1A210410811). 
We are grateful to the UK Materials and Molecular Modelling Hub for computational resources used in the research discussed in this review, which is partially funded by EPSRC (EP/P020194/1).

\bibliography{ref.bib}

\end{document}